\documentclass[12pt,a4paper]{article}

\usepackage[T1]{fontenc}
\usepackage[utf8]{inputenc}
\usepackage{lmodern}
\usepackage{microtype}
\usepackage[margin=1in]{geometry}
\usepackage{setspace}
\usepackage{amsmath,amssymb,amsfonts,amsthm}
\usepackage{mathtools}
\usepackage{bm}
\usepackage{siunitx}

\usepackage{booktabs}
\usepackage{array}
\usepackage{longtable}
\usepackage{tabularx}

\usepackage{graphicx}
\usepackage{float}
\graphicspath{{./}}
\usepackage{tikz}
\usetikzlibrary{positioning,arrows.meta,shapes.geometric,fit,calc,backgrounds,decorations.pathmorphing}

\usepackage{xcolor}
\definecolor{tealAccent}{RGB}{0,0,0}
\definecolor{tealBg}{RGB}{255,255,255}
\definecolor{navy}{RGB}{0,0,0}
\definecolor{amberAccent}{RGB}{0,0,0}
\definecolor{amberBg}{RGB}{255,255,255}

  {\par\medskip\noindent\textbf{Prediction \textnumero\,#1 - #2.}\par\nobreak\vspace{2pt}\ignorespaces}%
  {\par\medskip}

\newenvironment{insightbox}[1][Note]%
  {\par\medskip\noindent\textbf{#1:}\par\nobreak\vspace{2pt}\itshape\noindent\ignorespaces}%
  {\par\upshape\medskip}

  {\par\medskip\noindent\textbf{#1:}\par\nobreak\vspace{2pt}\ignorespaces}%
  {\par\medskip}

\usepackage[numbers,sort&compress]{natbib}

\usepackage[hidelinks]{hyperref}
\usepackage[noabbrev,capitalise]{cleveref}

\usepackage{titlesec}
\titleformat{\section}{\Large\bfseries}{}{0pt}{}
\titleformat{\subsection}{\large\bfseries}{}{0pt}{}
\titleformat{\subsubsection}{\normalsize\bfseries}{}{0pt}{}

\title{\textbf{Control Laws in Aging and Longevity}\\[4pt]
\large A Control Theory of Aging for Gerotherapeutic Drug Discovery:\\
State Space, Vector Fields, Modalities, Safety,\\
and the Minimum Safe Cost of Functional Restoration}

\author{%
Alex Zhavoronkov, PhD\textsuperscript{1,2,3,4,5},
Bud Mishra, PhD\textsuperscript{6}\\[4pt]
\small \textsuperscript{1}Insilico Medicine Hong Kong Ltd., Hong Kong SAR, China\\
\small \textsuperscript{2}Insilico Medicine Shanghai Ltd., Shanghai, China\\
\small \textsuperscript{3}Insilico Medicine AI Ltd., Masdar City, Abu Dhabi, UAE\\
\small \textsuperscript{4}Insilico Medicine US Inc., Cambridge, MA, USA\\
\small \textsuperscript{5}Buck Institute for Research on Aging, Novato, CA, USA\\
\small \textsuperscript{6}Courant Institute of Mathematical Sciences, New York University, New York, NY, USA\\[2pt]
\small Correspondence: \texttt{alex@insilicomedicine.com}
}

\date{\today}

\newcommand{\Vsafe}{\mathcal{U}_{\mathrm{safe}}}
\newcommand{\Vset}{\mathcal{V}}
\newcommand{\Vkernel}{\mathcal{K}}
\newcommand{\Rsafe}{\mathcal{R}_{\mathrm{safe}}}
\newcommand{\BAc}{\mathrm{BA}_{\mathrm{control}}}
\DeclareMathOperator*{\argmin}{arg\,min}

\begin{document}
\maketitle


\vspace{0.5em}
\noindent\textbf{Running title:} Control-theoretic framework for aging drug discovery.

\vspace{0.5em}
\noindent\textbf{Target journal:} \emph{Aging} (Impact Journals).

\vspace{0.5em}
\noindent\textbf{Note:} An expanded Supplementary Information accompanies this paper. It contains the full theory-family review, the complete state-space and observation model, the network-controllability and irreversibility extensions, the full worked-example parameterization and analytic derivations, the detailed implementation architecture, three translational case studies, the validation and power-analysis plan, the twenty predictions with full justifications, the complete empirical-scoring methodology, and connections to adjacent formal methods. Cross-references of the form S1, S2, and so on point to that document.

\clearpage

\section*{Abstract}\label{sec:abstract}\addcontentsline{toc}{section}{Abstract}
Existing aging theories describe what changes with age but do not prescribe how to intervene. We propose a control-theoretic framework that is not merely descriptive but prescriptive: it specifies which intervention, at which dose and sequence, under which safety constraints, will restore a measured biological state to a functional region. Aging is defined as progressive loss of safe controllability; biological age is the minimum safe control cost of functional restoration. Drugs are modeled as vector fields on biological state space whose non-commutativity, quantified by Lie brackets, predicts that intervention order determines outcome. The core differentiation from prior theories is operational: the framework outputs ranked targets, optimal sequences, safety-constrained protocols, and falsifiable predictions directly usable in drug discovery, rather than mechanistic ontologies or correlative biomarkers. We present a five-dimensional ODE model with analytic Lie-bracket derivation, a modality-aware control layer, three translational case studies, an implementation architecture with power analysis, and empirical scoring of aging interventions across five biological epochs. Twenty falsifiable predictions are enumerated. The central claim is that control-value reduction predicts translational success better than Hallmark annotation or biomarker reversal alone. If validated, this provides the missing interventional layer connecting aging biology to rational gerotherapeutic discovery.

\clearpage

\section{Introduction: From Mechanisms to Intervention Laws}\label{sec:introduction}
Despite decades of intensive research, aging biology lacks consensus on its most fundamental questions. A recent large-scale survey of leading experts revealed significant disagreement on the definition of aging, its causes, its onset, the meaning of rejuvenation, and whether aging should be classified as a disease \citep{gladyshev2024disagreement}, with none of these foundational questions receiving a majority opinion. The disagreement is not merely philosophical: different definitions lead to divergent experimental approaches, therapeutic strategies, and measures of success. The absence of a unifying quantitative framework that can accommodate these perspectives while generating falsifiable, intervention-relevant predictions motivates the present work.

The field has developed complementary frameworks. Evolutionary theories explain why natural selection may permit late-life decline \citep{medawar1952,williams1957,hamilton1966}. Damage and maintenance theories describe accumulation of molecular and cellular lesions \citep{harman1956,kirkwood1977}. The Hallmarks of Aging organize diverse mechanisms into a widely used ontology \citep{lopezotin2013,lopezotin2023}. SENS provides a repair-oriented catalogue of damage classes \citep{degrey2002}. Information theories emphasize epigenetic drift and partial reversibility through reprogramming \citep{lu2020,yang2023}. Geroscience frames aging mechanisms as modifiable drivers of chronic disease \citep{kennedy2014}. Hyperfunction theory highlights persistent growth signaling as a driver of late-life pathology \citep{blagosklonny2006}. These frameworks have transformed the field. As aging biology becomes increasingly connected to drug discovery, however, a different kind of theoretical object is required: it is no longer sufficient to ask only \emph{which} processes change with age. A therapeutic science of aging must answer a more operational question.

\begin{insightbox}[Operational question]
Given a measured biological state $x$, which intervention $u$, at which dose and duration, in which sequence, under which safety constraints, will move the system into a healthier functional region with acceptable risk and preserved future controllability?
\end{insightbox}

This question is not answered by a list of Hallmarks, a catalogue of damage classes, or a metaphor of information loss. Those frameworks identify relevant biology, but they do not by themselves specify the dynamics of the aged system, the state-dependent response to intervention, the safety-constrained set of admissible actions, or the objective function that defines therapeutic success. The relationship is analogous to that between thermodynamics and statistical mechanics: a phenomenological ontology organizes observations, while a dynamical and interventional layer supplies equations of motion, explicit inputs, response operators, constraints, objective functions, and predictions that can fail (Supplementary S1). The purpose of this paper is to propose that layer. We do not argue that existing theories are wrong; the framework depends on their insights. Our contribution is narrower but operationally important: we formulate aging as a problem of safe controllability and define biological age as the minimum safe control cost required to maintain or restore function under a specified intervention library.

Control theory is the natural mathematical language for this problem \citep{bellman1957,kalman1960,kirk1970,sontag1998}. Its central objects, including state variables, dynamics, inputs, response operators, reachable sets, value functions, constraints, and optimal policies, map directly onto the needs of interventional biology. A biological system has internal state variables that evolve under endogenous dynamics and stochastic perturbations; drugs, cell therapies, genetic perturbations, nutritional interventions, exercise, and surgery act as inputs whose effects are state-dependent, so the same intervention can be beneficial, neutral, or harmful depending on biological context. Therapeutic success is not movement toward an abstract youthful ideal but restoration or maintenance of function under acceptable risk and preserved future controllability. In this framework a drug is a vector field on biological state space, a state-dependent transformation operator that pushes the system along particular directions. A target is valuable if its modulation reduces the optimal control cost of returning the system to a functional viability set; a combination is valuable if its component vector fields expand the reachable safe set more than either intervention alone; and a sequence matters when the vector fields do not commute.

This last point provides the framework's central geometric content. Local controllability, in the sense of \citet{kalman1960,kalman1963} and the geometric control theory of \citet{nijmeijer1990} and \citet{isidori1995}, can be improved by exploiting the noncommutativity of pairs of interventions. Given intervention vector fields $g_{A}$ and $g_{B}$, sequential application along $g_{A}$ then $g_{B}$ then their reverses generically returns the system to a state displaced from the start by an amount proportional to $\varepsilon^{2}$ along a new direction, quantified by the Lie bracket $[g_{A},g_{B}](x) = Dg_{B}(x)\,g_{A}(x) - Dg_{A}(x)\,g_{B}(x)$. The same logic distinguishes \emph{local} controllability (a small reachable set near the current state) from \emph{global} controllability (reachability of any youthful state from the aged state). An aged tissue may retain local controllability while having lost global controllability because irreversible coordinates have collapsed the reachable set. The progressive collapse from global to merely local controllability, and eventually to loss of even local controllability in critical dimensions, \emph{is} the aging process under this framework. A clinical running illustration (type 2 diabetes monitored by continuous glucose monitoring) is developed in Supplementary S2 and referenced where relevant below.

Three developments make this framework timely. First, multi-omics and single-cell technologies increasingly permit high-dimensional measurement of biological state. Second, perturbational datasets, including CRISPR screens, LINCS/CMap profiles \citep{lamb2006,subramanian2017}, drug-response atlases, and animal intervention studies, provide empirical information about how systems respond to inputs. Third, machine-learning methods can in principle estimate state representations, infer dynamics, rank targets, model response heterogeneity, and integrate heterogeneous data \citep{zhavoronkov2019a,zhavoronkov2019b,zhavoronkov2019c}. These technologies do not eliminate the need for theory; they make a mathematically explicit theory of interventional aging tractable. This paper presents the framework and a computational pipeline for prioritizing gerotherapeutic interventions; it does not report a trained engine that has learned intervention vector fields from perturbational data. The estimation of state-dependent vector fields from data is developed as a method and flagged as future and companion work (Supplementary S5), not as an existing capability. We have previously argued that artificial intelligence provides machinery to integrate heterogeneous aging data into longevity pipelines \citep{zhavoronkov2019ai_aging}, demonstrated an early pathway-activation instantiation of aging as a multi-dimensional problem \citep{zhavoronkov2014pathway}, and shown that aging clocks can serve as instruments for therapeutic target discovery \citep{chen2025clock}. The control-theoretic framework supplies the mathematical scaffold connecting these advances: it defines the state space in which clocks operate, the objective function that target discovery optimizes, and the safety constraints under which such interventions must be evaluated. Its value will be measured by whether it improves prediction, experimental design, and therapeutic outcomes.

\section{Background: The Landscape of Theories of Aging}\label{sec:background}
Aging is a hierarchy of phenomena spanning molecular damage, cellular dysregulation, tissue remodeling, organ decline, and rising mortality risk, and the major theoretical families address different levels of it: evolutionary \citep{medawar1952,williams1957,hamilton1966,kirkwood1977,kirkwood1979}, damage and repair \citep{harman1956,harman1972,wallace2005,degrey2002,degrey2007}, the Hallmarks ontology \citep{lopezotin2013,lopezotin2023}, information and reprogramming \citep{sinclair2019,yang2023,ocampo2016,lu2020,gill2022}, geroscience \citep{kennedy2014,sierra2017,barzilai2018}, hyperfunction \citep{blagosklonny2006,blagosklonny2013,harrison2009,miller2011,lamming2013,kennedy2016}, and reliability and resilience \citep{gavrilov1991,gompertz1825,scheffer2009,scheffer2012,gao2016}. Each contributes essential insight; none alone provides a full control law for intervention (a detailed per-theory review is given in Supplementary S1). Despite their differences, most lack eight components required for drug discovery: (1)~\textbf{state variables}, the mathematical state of the system; (2)~\textbf{dynamics}, how the state evolves without intervention; (3)~\textbf{control inputs}, what can be manipulated; (4)~\textbf{response operators}, how each intervention affects the state and how that effect depends on context; (5)~\textbf{constraints}, which interventions are unsafe, infeasible, or toxic; (6)~\textbf{objective functions}, what counts as improvement; (7)~\textbf{optimality conditions}, which intervention or sequence minimizes cost while restoring function; and (8)~\textbf{falsifiable predictions}, what quantitative outcomes would refute the framework. Their absence does not make previous theories wrong; it indicates they were built to answer different questions. Drug discovery requires a theory of safe control, which the next section formalizes.

\section{The Control Theory of Aging}\label{sec:framework}

\subsection{Biological state space}\label{sec:statespace}
We represent a biological system as a latent state vector $x(t)\in\mathbb{R}^{n}$ describing a cell population, tissue, organ, physiological subsystem, or organism. It is latent because no platform measures all relevant variables; observed data are noisy projections,
\begin{equation}
  y(t) \;=\; h\!\bigl(x(t)\bigr) + \varepsilon(t),
  \label{eq:observation}
\end{equation}
where $y(t)$ may include transcriptomic, proteomic, metabolomic, epigenetic, imaging, clinical, and functional measurements; $h$ is an observation function; and $\varepsilon(t)$ represents measurement error and unobserved variation. The components of $x(t)$ need not correspond one-to-one to measured biomarkers; in practice $x(t)$ is inferred as a structured representation learned from multi-omics, clinical, and perturbational data, with coordinates spanning epigenetic, transcriptional, metabolic, mitochondrial, proteostatic \citep{hipp2019}, senescent, immune, regenerative, matrix, vascular, and systemic-functional axes. The full specification, including hierarchical state representations and measurement specifics, is given in Supplementary S2.

\subsection{The functional viability set}\label{sec:viability}
Aging research often describes intervention goals as ``rejuvenation'' or reversal of biological age. These are imprecise: the goal of medicine is not to make every molecular feature identical to youth but to preserve or restore function safely. We define a functional \textbf{viability set}
\begin{equation}
  \Vset \;=\; \bigl\{\, x \,:\, F_{i}(x) \ge \theta_{i},\; \forall\,i \,\bigr\},
  \label{eq:viabilityset}
\end{equation}
where each $F_{i}(x)$ is a functional measure and $\theta_{i}$ an acceptable threshold (cardiac reserve, renal filtration, immune competence, wound healing, muscle force, cognition, pulmonary compliance, hematopoietic output, metabolic flexibility, vascular reactivity). Thresholds need not be those of a young adult; they may be age-appropriate, disease-specific, or patient-specific. The viability set avoids treating youth as an attractor: developmental trajectories are not simply reversible aging trajectories, and a function-based objective is more biologically and clinically appropriate.

The separable form in \Cref{eq:viabilityset} treats each functional measure independently, but physiological constraints are frequently interdependent. The general form of the viability set is
\begin{equation}
  \Vset \;=\; \bigl\{\, x \,:\, G(x) \ge 0 \,\bigr\},
  \label{eq:viabilityset-general}
\end{equation}
where $G:\mathbb{R}^{n}\to\mathbb{R}^{p}$ is a vector of joint constraint functions and the inequality is taken componentwise. The separable threshold form $F_{i}(x)\ge\theta_{i}$ is the special case $G_{i}(x)=F_{i}(x)-\theta_{i}$ in which each constraint involves a single functional coordinate. Joint constraints capture the common situation in which an acceptable value of one function depends on other coordinates: an acceptable body weight depends jointly on height, age, and sex; an acceptable cardiac output depends on body size and metabolic demand; an acceptable hematocrit depends on altitude and sex. When such dependencies matter, $G$ is specified directly rather than assembled from independent thresholds, and the constructions below carry over unchanged with $\Vset$ read in its general form.

\begin{figure}[t]
\centering
\includegraphics[width=\textwidth,keepaspectratio]{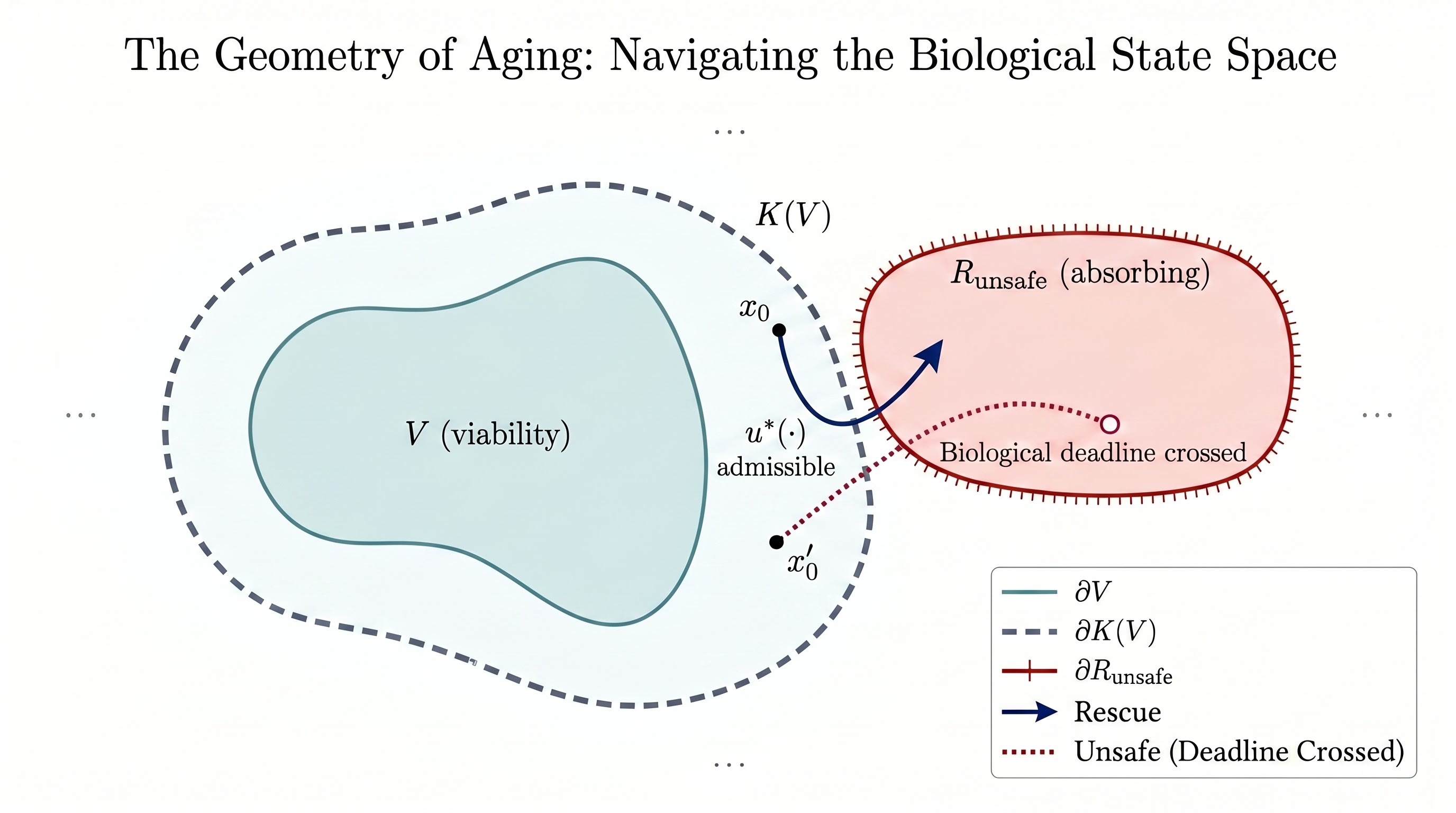}
\caption{The geometry of aging: navigating the biological state space. A 2D projection of the $n$-dimensional biological state space showing the viability set $\mathcal{V}$ (region of acceptable function), the viability kernel $\mathcal{K}(\mathcal{V})$ (recovery zone), and the absorbing region $\mathcal{R}_{\mathrm{unsafe}}$ (irreversible loss). Consistent with the finite-entry-time definition in \Cref{eq:kernel}, the kernel contains $\mathcal{V}$ and also states that begin outside $\mathcal{V}$ from which an admissible safe control reaches $\mathcal{V}$ within a finite entry time and remains thereafter, so $\mathcal{V}\subseteq\mathcal{K}(\mathcal{V})$. Trajectory~A (rescue) begins outside $\mathcal{V}$ but inside $\mathcal{K}(\mathcal{V})$ and returns to $\mathcal{V}$ under optimal intervention $u^{\ast}(\cdot)$; trajectory~B begins outside $\mathcal{K}(\mathcal{V})$, crosses the biological deadline, and terminates in $\mathcal{R}_{\mathrm{unsafe}}$.}
\label{fig:viability}
\end{figure}

Before defining the viability kernel we specify the admissible safe control set $\Vsafe$, the set of control functions $u(\cdot)$ satisfying the safety, feasibility, pharmacologic, dosing, and ethical constraints of the problem, including the hard and soft constraints developed in \Cref{sec:ctrl-loss}; $\Vsafe$ is in general state-dependent, so the same intervention may be admissible in one state and forbidden in another. A related concept is then the \textbf{viability kernel}, the set of initial states from which some admissible safe control keeps the system in $\Vset$, or drives it into $\Vset$ within a finite entry time and holds it there over the horizon,
\begin{equation}
  \Vkernel \;=\; \bigl\{\, x_{0} \,:\, \exists\,u(\cdot)\in\Vsafe \text{ and } t_{e}\in[0,T] \text{ such that } x_{t}\in\Vset \text{ for all } t\in[t_{e},T] \,\bigr\}.
  \label{eq:kernel}
\end{equation}
This definition admits initial states that lie outside $\Vset$ but can be steered into it and maintained thereafter, so that $\Vset\subseteq\Vkernel$: any state already in $\Vset$ that can be held there has entry time $t_{e}=0$. In the classical viability-theory sense the viability kernel is a subset of the constraint set; the object in \Cref{eq:kernel} is the corresponding \emph{reach-avoid} (capture-basin) generalization that additionally admits states outside $\Vset$ from which $\Vset$ can be safely reached and thereafter maintained, and we use it throughout in this sense. States outside the kernel may still be transiently improved but cannot be brought into and held within functional bounds under available safe interventions, providing a formal language for frailty, irreversibility, and late-stage disease (\Cref{fig:viability}).

The viability kernel generalizes the classical basin of attraction. For an autonomous dynamical system $\dot{x}=f(x)$ with a stable attractor $\Vset$, the basin of attraction is the set of initial states whose uncontrolled trajectories converge to $\Vset$ as $t\to\infty$. The viability kernel differs in three respects. It is control-dependent: membership requires the existence of an admissible control, not spontaneous convergence, so the kernel expands as the intervention library grows and contracts as it shrinks. It is safety-constrained: only controls in $\Vsafe$ are allowed, so states reachable only through forbidden trajectories are excluded. It is finite-horizon: entry and maintenance are required over $[0,T]$ rather than asymptotically, which matters when biological deadlines make the asymptotic limit irrelevant. When the control set is empty and $T\to\infty$, the kernel reduces to the ordinary basin of attraction of $\Vset$.

\subsection{Aging dynamics without intervention}\label{sec:aging-dynamics}
In the absence of intervention, the biological state evolves according to stochastic dynamics,
\begin{equation}
  dx_{t} \;=\; f\!\bigl(x_{t},a,e,g\bigr)\,dt \;+\; \Sigma(x_{t})\,dW_{t},
  \label{eq:uncontrolled-sde}
\end{equation}
where $f$ is the endogenous drift; $a$ chronological age; $e$ environment; $g$ genotype; $\Sigma(x_{t})$ the state-dependent noise matrix; and $W_{t}$ a Wiener process. The inclusion of $a$ does not imply that age is an independent causal force; it indexes time-dependent exposures, cumulative damage, and changing regulatory regimes. The noise matrix is state-dependent because youthful well-buffered states have small $\Sigma$ while pre-frail and senescent-cell-rich states show inflated variance; estimation procedures are detailed in Supplementary S2. Aging corresponds to several changes: the drift increasingly moves the system away from $\Vset$; stochastic instability rises with transcriptional noise, epigenetic drift, and immune contraction \citep{bahar2006,enge2017,martinez2017}; recovery after perturbation slows, a known early-warning signal near critical transitions \citep{scheffer2009,scheffer2012}; the dimensionality of safe control decreases; and the cost of restoration increases until, at advanced stages, no safe path may exist. Gompertz-like increases in hazard \citep{gompertz1825} may emerge as the probability of crossing viability boundaries rises with accumulating drift, noise, and loss of redundancy.

\subsection{Intervention as control}\label{sec:intervention}
Interventions enter the dynamics as control inputs,
\begin{align}
  dx_{t} &\;=\; f(x_{t})\,dt \;+\; \sum_{j=1}^{m} g_{j}(x_{t})\,u_{j}(t)\,dt \;+\; \Sigma(x_{t})\,dW_{t} \label{eq:controlled-sde-sum}\\
  &\;=\; f(x_{t})\,dt \;+\; G(x_{t})\,u_{t}\,dt \;+\; \Sigma(x_{t})\,dW_{t}, \label{eq:controlled-sde-matrix}
\end{align}
where $u_{j}(t)$ denotes the intensity, dose, schedule, or exposure of intervention $j$, $g_{j}(x)$ is the corresponding intervention vector field, and $G(x)$ collects all available fields. \Cref{eq:controlled-sde-matrix} is the core translation from aging biology to drug discovery: a drug is a transformation of biological state rather than a label or pathway inhibitor. The vector field $g_{j}(x)$ is state-dependent, which is biologically central. Rapamycin in a metabolically overactive, inflammatory, pre-frail state may reduce pathology; in a frail state requiring wound repair it may impair recovery \citep{demaria2014,ritschka2017,lopezotin2016}. The sign of an intervention is therefore not intrinsic but depends on the value function: locally, intervention $j$ is beneficial when $g_{j}(x)^{\top}\nabla V(x) < 0$ and harmful when $g_{j}(x)^{\top}\nabla V(x) > 0$, formalizing context dependence as a geometric property rather than a rhetorical caveat.

\subsection{Biological age as control cost: the core definition}\label{sec:ba-cost}
We define the optimal control value function
\begin{equation}
\boxed{\;
V(x_{0},T) \;=\; \min_{u(\cdot)\in\Vsafe}\; \mathbb{E}\!\left[\;\int_{0}^{T}\Bigl(\ell\bigl(x_{t}\bigr) + \lambda\,c\bigl(u_{t}\bigr) + \rho\,r\bigl(x_{t},u_{t}\bigr)\Bigr)\,dt \;+\; \Phi\bigl(x_{T}\bigr)\;\right]
\;}
\label{eq:valuefunction}
\end{equation}
subject to
\begin{equation}
dx_{t} = f(x_{t})\,dt + G(x_{t})\,u_{t}\,dt + \Sigma(x_{t})\,dW_{t}.
\label{eq:subjectto}
\end{equation}
Here $\ell(x_{t})$ penalizes functional loss or distance from $\Vset$; $c(u_{t})$ penalizes intervention burden; $r(x_{t},u_{t})$ penalizes toxicity and safety risk; $\Phi(x_{T})$ penalizes terminal dysfunction; and $\Vsafe$ is the admissible safe control set defined in \Cref{sec:viability}. A simple choice is $\ell(x)=d(x,\Vset)$ or $\ell(x)=\sum_{i}w_{i}\,\max(0,\theta_{i}-F_{i}(x))$. We define \textbf{control biological age} as

\begin{equation}
\BAc(x_{0}) \;=\; \varphi\!\bigl(V(x_{0},T)\bigr),
\label{eq:bacontrol}
\end{equation}
where $\varphi$ maps restoration cost to a clinically interpretable scale. Crucially, $\BAc$ is \emph{intervention-relative}: $V$ depends on the admissible intervention set $\mathcal{U}$, measurement model $\mathcal{M}$, cost function $\mathcal{C}$, horizon $T$, and safety constraints. A state may be highly controllable using cell therapy and gene editing but poorly controllable using only small molecules.

\begin{insightbox}[Core definition]
A young system has low $V$ and returns to function through endogenous repair or modest intervention. A middle-aged system has higher $V$, requiring stronger intervention. A frail system has high $V$, reflecting narrow safety margins. A terminally damaged system has effectively infinite $V$: no admissible control path reaches $\Vset$. Chronological age measures time lived; biomarker age \citep{hannum2013,horvath2013,levine2018,lu2019} predicts age or outcomes from molecular features; Hallmark burden scores age-associated processes; control biological age measures how difficult it is to restore or maintain function safely. Only the last is directly actionable.
\end{insightbox}

Around a state or trajectory, dynamics can be approximated as $\dot{x}=A(a)\,x + B(a)\,u$, with $A(a)$ age-dependent endogenous dynamics and $B(a)$ intervention susceptibility. The controllability Gramian over horizon $T$ is
\begin{equation}
  W(T,a) \;=\; \int_{0}^{T} e^{A(a)\tau}\,B(a)\,B(a)^{\top}\,e^{A(a)^{\top}\tau}\,d\tau,
  \label{eq:gramian}
\end{equation}
and for linear systems the minimum energy to move from $x_{0}$ to $x_{T}$ is
\begin{equation}
  E^{*}(x_{0},x_{T},a) \;=\; \bigl(x_{T} - e^{A(a)T}x_{0}\bigr)^{\top} W(T,a)^{-1} \bigl(x_{T} - e^{A(a)T}x_{0}\bigr).
  \label{eq:minenergy}
\end{equation}
With age, the smallest eigenvalues of $W$ should decline and the minimum restoration energy should increase. The value-function biological age connects to classical survival analysis \citep{cox1972,aalen1978,kalbfleisch2002}: with hazard $\lambda(t\mid x(t))$ and survival $S(t)=\exp(-\int_{0}^{t}\lambda(s\mid x(s))\,ds)$, the hazard is a monotone function of the value function,
\begin{equation}
  \lambda\bigl(t \mid x(t)\bigr) \;=\; \Psi\!\bigl(V(x(t),T_{\mathrm{ref}})\bigr),
  \label{eq:hazard-V}
\end{equation}
for a non-negative link $\Psi$ and reference horizon $T_{\mathrm{ref}}$. This formalizes biological-age acceleration ($dV/dt$ above the cohort mean), geroprotection (kernel expansion that shifts $V$ downward), and mortality-rate deceleration (survivor selection for low-$V$ states). The horizon $T$ is a problem-specification choice (trial duration for trials; $T_{\mathrm{ref}}=5$ years for cross-population comparison) that shortens as biological deadlines approach; the full treatment of horizon selection and the Kalman controllability/observability decomposition appears in Supplementary S2.

\subsection{Network controllability, irreversibility, and biological deadlines}\label{sec:network-control}
\begin{figure}[t]
\centering
\includegraphics[width=\textwidth,keepaspectratio]{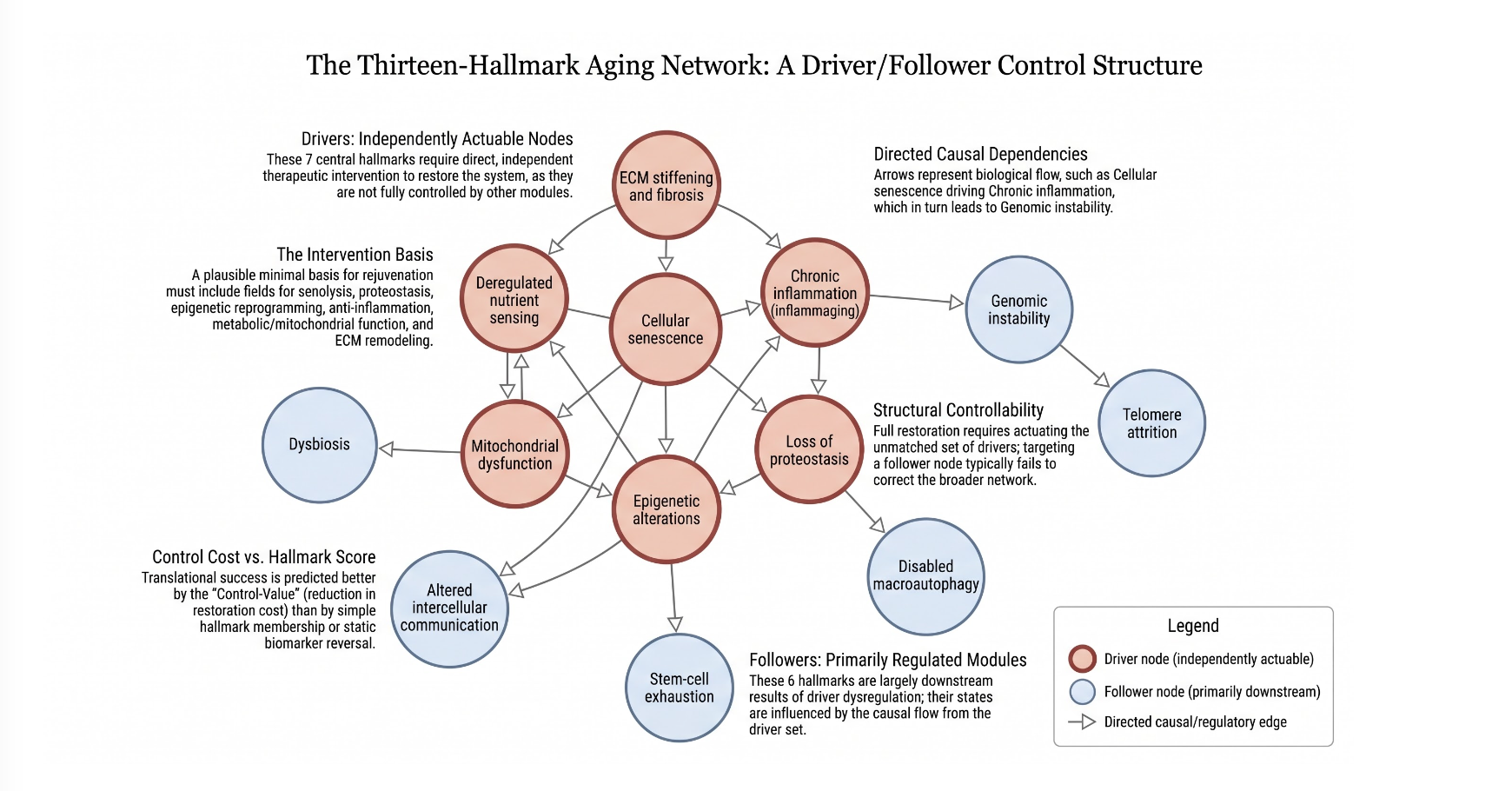}
\caption{Thirteen-hallmark network of aging with driver and follower classification. Drivers (cellular senescence, loss of proteostasis, deregulated nutrient sensing, epigenetic alterations, chronic inflammation, mitochondrial dysfunction, ECM stiffening and fibrosis) have clinically tractable actuating modality classes; followers are primarily downstream. Edges encode known causal and regulatory dependencies.}
\label{fig:hallmark-network}
\end{figure}

The framework is closely related to, but distinct from, the network-controllability program of \citet{liu2011}, in which a linearized network $\dot{x}=Ax+Bu$ is structurally controllable when a minimum set of driver nodes determined by maximum matching is actuated. In aging, nodes represent senescent burden, mitochondrial dysfunction, proteostatic stress, stem-cell exhaustion, epigenetic drift, inflammation, matrix remodeling, nutrient sensing, telomere and genomic state, intercellular communication, and autophagy, with directed edges encoding causal dependencies (\Cref{fig:hallmark-network}). This clarifies why single-target geroprotectors have limited restorative power: if unmatched nodes are distributed across hallmark modules, one driver is insufficient, so effective late-life rejuvenation will usually require multiple independent intervention vector fields. Our framework extends this program in four ways, detailed in Supplementary S3: we ask whether control is achievable at acceptable biological cost rather than only in principle; we include safety constraints through the viability-constrained reachable set
\begin{equation}
  \Rsafe(x_{0},T) \;=\; \bigl\{\, x_{T} \,:\, \exists\,u(\cdot)\in\Vsafe,\; x(0)=x_{0},\; x(T)=x_{T},\; x(t)\in\Vset\;\forall t\in[0,T]\,\bigr\};
  \label{eq:reachable-safe}
\end{equation}
we treat aging as time-varying, since the same drug field can have different effects across ages; and we replace binary controllability with the continuous value function. A plausible minimal intervention basis includes anti-senescent, proteostasis/nutrient-sensing, epigenetic-reprogramming, anti-inflammatory, mitochondrial/metabolic, and matrix/fibrosis fields, though whether it suffices is an empirical network-identifiability problem.

Aging is not composed solely of reversible deviations. We decompose the state into reversible and irreversible coordinates, $x(t)=[x_{R}(t),x_{I}(t)]$, where many components of $x_{I}$ (somatic driver mutations, mitochondrial DNA deletions, neuronal, nephron, and oocyte loss, severe matrix crosslinking, architectural fibrosis, clonal expansion, thymic stromal collapse) are monotonic or nearly monotonic under the admissible control set, reflecting the clinically available controls rather than physical impossibility. An \textbf{absorbing boundary} is a region beyond which return to $\Vset$ is impossible under the specified library and safety constraints: with $\mathcal{R}(x_{0},T,\Vsafe)$ the safe-reachable set, $x_{0}$ is controllable to viability when $\mathcal{R}(x_{0},T,\Vsafe)\cap\Vset\ne\emptyset$, and an absorbing boundary is reached when this intersection is empty. Absorbing boundaries imply \textbf{biological deadlines}, time points after which the optimal policy changes from restoration to compensation, replacement, or palliation,
\begin{equation}
  t^{*} \;=\; \sup\!\left\{\, t \,:\, \Pr\!\bigl[\,\mathcal{R}(x(t),T,\Vsafe)\cap\Vset\ne\emptyset\,\bigr] \ge 1-\alpha \,\right\}.
  \label{eq:deadline}
\end{equation}
This matters for preventive geroscience: late rescue is biologically more difficult and clinically more dangerous than early stabilization. When a boundary is crossed, the admissible set can sometimes be expanded by \emph{prosthetic controls} (insulin pumps, L-DOPA, dialysis, cochlear implants) that compensate for an irreversible coordinate without restoring its biological function, yielding a therapeutic hierarchy: prevent boundary crossing, restore biologically, compensate prosthetically, or manage decline. Extended material on the rare-event (Luria and Delbr\"uck) statistics of irreversible coordinates \citep{luria1943,jee2016}, phase transitions and tipping points, propagation of irreversible losses, and a worked hippocampal example appears in Supplementary S3.

\section{Drugs as Vector Fields}\label{sec:drugs}

\subsection{The drug-as-vector-field concept}\label{sec:vf-concept}
Drug discovery often describes compounds by their primary target, compressing a high-dimensional intervention into a scalar mechanism. In the control-theoretic framework, a drug induces a vector field $g_{j}(x)$ on biological state space whose effect depends on current state, dose, duration, tissue context, pharmacokinetics, pharmacodynamics, and endogenous feedback. This provides a natural language for responder heterogeneity (two individuals of the same chronological age but different $x$ may experience opposite effects) and for dose-response, since optimal dosing is selection of $u(t)$ that minimizes the value function under constraints rather than maximization of target engagement. The drug-as-vector-field is the mean-field projection of the more general drug-as-distribution operator $\mu_{t+\Delta t}=T_{u}\mu_{t}$ used in single-cell perturbation modeling \citep{lotfollahi2023}; the two agree under weak heterogeneity and diverge when subpopulation selection or tail outcomes dominate, in which case the distributional lift is required (Supplementary S4).

\subsection{Non-commutativity: why order matters}\label{sec:non-commute}
Control theory predicts that sequence can matter independently of the components. Two vector fields commute only if their Lie bracket is zero,
\begin{equation}
  [g_{A},g_{B}](x) \;=\; \frac{\partial g_{B}}{\partial x}(x)\,g_{A}(x) \;-\; \frac{\partial g_{A}}{\partial x}(x)\,g_{B}(x).
  \label{eq:liebracket-def}
\end{equation}
When $[g_{A},g_{B}]\ne 0$, the order changes the trajectory: $A\to B\ne B\to A$. A senolytic may remove cells that constrain tissue remodeling and thereby change the response to reprogramming, whereas reprogramming before senolysis may act on a tissue still dominated by SASP, fibrosis, or immune dysfunction. The bracket has a precise operational meaning: applying $g_{A}$, then $g_{B}$, then $-g_{A}$, then $-g_{B}$, each for time $\varepsilon$, returns the system to its start to leading order, with a second-order residual displacement of exactly $\varepsilon^{2}\,[g_{A},g_{B}](x)+O(\varepsilon^{3})$. The bracket is therefore the new direction in state space exposed by exploiting noncommutativity, a degree of freedom inaccessible to either intervention alone. In the geometric control language of \citet{nijmeijer1990} and \citet{isidori1995}, the controllability Lie algebra at $x$ determines the dimension of the locally accessible set; interventions that fail to span this algebra cannot reach the youthful viability set even at unbounded duration and dose. The expected improvement from sequencing is quantified by $\|[g_{A},g_{B}]\|$ projected onto the gradient of $V$. These sequence effects are not generally predicted by Hallmark annotation, which can suggest combining senolysis and reprogramming but does not define when one should precede the other. \Cref{sec:worked-example} carries out the explicit analytic Lie-bracket calculation for senolysis and reprogramming.

\begin{insightbox}[Experimental Test]
For a pair of interventions $A$ and $B$, estimate $g_{A}$, $g_{B}$, and their Lie bracket in a relevant aged tissue model. Predict whether $A\to B$, $B\to A$, simultaneous treatment, or monotherapy best reduces $V$. Test predefined functional endpoints, safety markers, and durability. If strong predicted non-commutativity repeatedly fails to correspond to sequence-dependent outcomes, the model is wrong.
\end{insightbox}

\subsection{Controllability loss, modality classes, and safety as a first-class object}\label{sec:ctrl-loss}
Aging reduces controllability in several ways: intervention susceptibility declines (stem-cell exhaustion, vascular rarefaction, immune exhaustion); structural variables make the target set harder to reach (fibrosis, crosslinking, neuronal and nephron loss); safety constraints tighten with frailty, shrinking the admissible set; and stochasticity rises. Many interventions fail because they are applied after a controllability threshold has been crossed. The control input $u_{j}(t)$ is not restricted to small molecules: modern modalities (biologics, mRNA and lipid nanoparticles, viral gene delivery, genome and epigenome editing, engineered immune cells, stem-cell products, synthetic circuits) differ fundamentally in reversibility, latency, durability, targeting, titratability, immunogenicity, and monitoring. Each admissible control should be annotated as
\begin{equation}
  u_{j} = (m_{j},d_{j},\tau_{j},\rho_{j},\eta_{j},\sigma_{j}),
  \label{eq:modality-tuple}
\end{equation}
with modality class, dose, timing, reversibility, tissue-targeting profile, and uncertainty (\Cref{tab:modalities}). A practical policy follows a \textbf{reversibility hierarchy}: in early or uncertain states, reversible interventions are preferred to perturb, observe, and update the state estimate; only after the trajectory is validated should the controller escalate to durable modalities. The escalation rule makes a permanent control $u_{\mathrm{perm}}$ admissible only if
\begin{equation}
  \Pr\!\bigl[\,\Delta V(x;u_{\mathrm{perm}}) > \Delta V(x;u_{\mathrm{rev}}) + \epsilon\,\bigr] > \gamma
  \quad\text{and}\quad
  \Pr\!\bigl[\,x(t)\text{ forbidden}\mid u_{\mathrm{perm}}\,\bigr] < \alpha
  \label{eq:escalation}
\end{equation}
over a clinically meaningful horizon, preventing premature use of irreversible modalities under model uncertainty.

\begin{table}[t]
\centering
\footnotesize
\caption{Modality classes and control properties.}\label{tab:modalities}
\begin{tabularx}{\textwidth}{@{}l l l l X X@{}}
\toprule
\textbf{Modality} & \textbf{Reversibility} & \textbf{Onset} & \textbf{Duration} & \textbf{Principal safety concerns} & \textbf{Control implication}\\
\midrule
Small molecules & High & Hours-weeks & Short-medium & Off-target toxicity, interactions, infection, metabolic & Iterative MPC, dose titration, sequence exploration\\
Biologics / antibodies & Moderate & Days-weeks & Weeks-months & Immunogenicity, cytokine effects, target depletion & Sustained but adjustable; requires washout modeling\\
mRNA / LNP & Moderate-high & Hours-days & Days-weeks & Innate immune activation, reactogenicity, tissue accumulation & Pulsed expression; reversible program testing\\
AAV gene delivery & Low & Weeks & Months-years & Immunogenicity, insertional risk, no redose, persistent expression & High-confidence state selection, long-horizon safety, stopping rules\\
Base / prime editing & Very low & Days-weeks & Permanent & Off-target, clonal expansion, p53 selection, germline risk & Reserve for irreversible lesions with strong causal evidence\\
Epigenome editing & Variable & Days-weeks & Weeks-years & Identity loss, dedifferentiation, lineage mis-programming & Identity-preservation constraints, longitudinal monitoring\\
Engineered immune cells & Low-moderate & Days-weeks & Months-years & CRS, off-target killing, exhaustion, autoimmunity, transformation & Population dynamics, kill switches, antigen escape\\
Senolytic cell therapy & Low-moderate & Days-weeks & Durable depletion & Loss of beneficial senescent cells, impaired repair, inflammation & Use when senescent burden causal and localized\\
iPSC-derived cell therapy & Low & Weeks-months & Years & Teratoma, ectopic tissue, rejection, mis-integration & Replacement for crossed boundaries; release + surveillance\\
Synthetic circuits & Variable & Hours-weeks & Programmable & Circuit escape, mutation, immune recognition & Enables closed-loop control with fail-safes\\
\bottomrule
\end{tabularx}
\end{table}

Safety is not a secondary penalty appended after efficacy; it is the central constraint that defines the admissible control set. We define a forbidden region of unacceptable risk (excessive clonal hematopoiesis, loss of epithelial identity, oncogenic proliferation signatures, severe immunosuppression, cytokine-storm risk, impaired wound repair, fibrosis beyond threshold) and require an admissible policy to keep the probability of entering it below a tolerated level $\alpha$ over the horizon, with $\alpha$ much smaller for preventive aging than for life-threatening disease. \textbf{Hard constraints} (predicted malignant transformation, expansion of high-risk clones, severe immune collapse, persistent identity loss, organ toxicity above stopping thresholds, germline exposure) exclude a control regardless of predicted benefit; \textbf{soft constraints} (mild infection risk, transient wound-healing impairment, reversible cytopenias) enter as context-dependent penalties. The admissible set is state-dependent: the same intervention can be admissible in one patient and forbidden in another. To prevent \textbf{reward hacking}, where measured function improves while latent damage accumulates, the terminal cost includes future controllability,
\begin{equation}
  \Phi\bigl(x(T)\bigr) \;=\; \Phi_{\mathrm{function}}\!\bigl(x(T)\bigr) + \beta_{1}\,D_{\mathrm{irreversible}} + \beta_{2}\,R_{\mathrm{cancer}} + \beta_{3}\,R_{\mathrm{immune}} + \beta_{4}\,C_{\mathrm{future}}^{-1},
  \label{eq:terminal-safety}
\end{equation}
so that an intervention is successful only if it preserves or expands the future safe reachable set. The relationship to classical PK/PD and systems pharmacology \citep{holford1981,bhalla1999,sorger2011,lamb2006,subramanian2017}, and the full treatment of modality reversibility, hard and soft constraints, and reward-hacking prevention, are developed in Supplementary S4.

\section{Worked Computational Example: Aged Murine Liver}\label{sec:worked-example}
To make the framework concrete we use a deliberately low-dimensional, fully specified model of aged murine liver. The purpose is not to claim that five variables capture hepatic aging, but to show how hallmarks, interventions, safety constraints, non-commutativity, and value-function scoring can be placed into a single computable system simple enough to be reproduced, falsified, and extended. Let $x(t)=(s,d,e,r,f)\in[0,1]^{5}$ collect senescent-cell fraction $s$, oxidative damage $d$, epigenetic information integrity $e$, regenerative capacity $r$, and fibrotic burden $f$, initialized at an aged state $x_{0}=(0.15,0.40,0.55,0.40,0.30)$. The controlled stochastic dynamics, literature-calibrated illustrative parameterization \citep{baker2011,xu2018,harrison2009,horvath2013,ocampo2016,lu2020}, the viability set, and the full equations are specified in Supplementary S5. Controls are $u_{\mathrm{sen}}$ (senolytic), $u_{\mathrm{rep}}$ (reprogramming), $u_{\mathrm{rap}}$ (rapamycin), and $u_{\mathrm{tnik}}$ (TNIK inhibitor), with senolytic courses and reprogramming pulses normalized to unit integral.

\begin{figure}[H]
\centering
\includegraphics[width=0.95\textwidth,keepaspectratio]{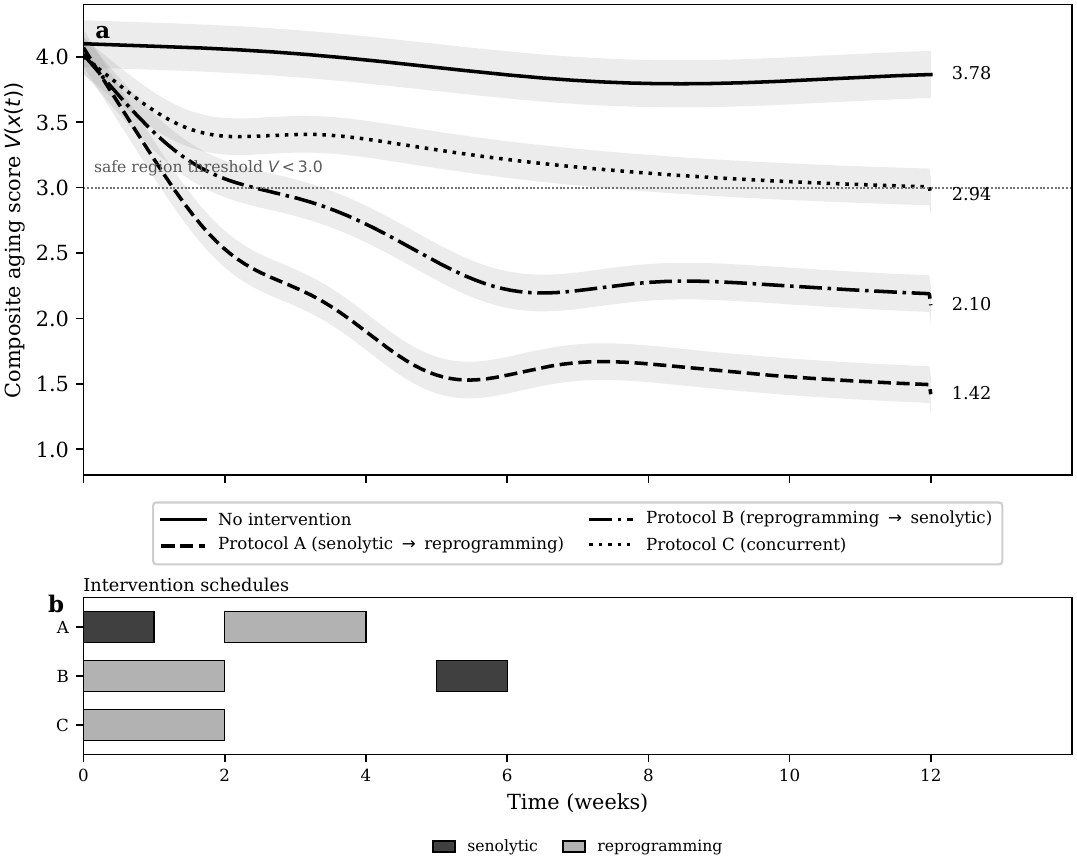}
\caption{(a) Worked-example terminal-cost trajectory for the intervention protocols in the five-variable aged murine liver model $x=(s,d,e,r,f)$ over the $T=56$-day horizon ($n=200$ stochastic realizations per protocol; explicit Euler and Maruyama SDE integration at $\Delta t=0.25$ days). Solid curve: no intervention. Dashed: Protocol~A (senolytic $\to$ reprogramming). Dash-dot: Protocol~B (reprogramming $\to$ senolytic). Dotted: Protocol~C (simultaneous senolytic $+$ reprogramming). Shaded bands are 95\% bootstrap confidence intervals. (b) Intervention schedules for the three protocols (dark bars: senolytic course; light bars: reprogramming pulses). Full equations, parameters, and protocol schedules are specified in Supplementary S5.}
\label{fig:worked-example}
\end{figure}

\begin{figure}[H]
\centering
\includegraphics[width=0.95\textwidth,keepaspectratio]{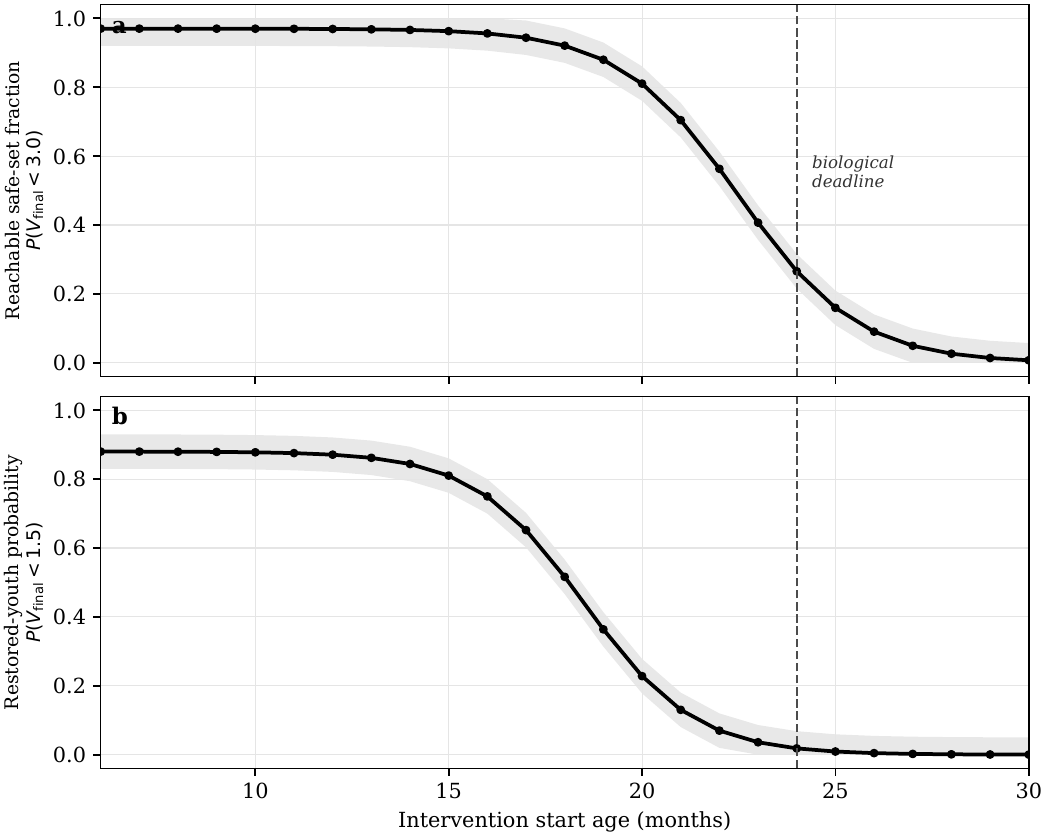}
\caption{Loss of controllability with age in the five-variable aged-liver model. Upper panel: reachable safe-set fraction (the probability that Protocol~A over the $T=56$-day horizon returns the system to a functionally safe region) as a function of intervention start age. Lower panel: restored-function success probability under a stricter viability threshold. Shaded bands are 95\% bootstrap confidence intervals over $n=300$ stochastic realizations per age point. The dashed vertical line marks the biological deadline beyond which no admissible protocol returns the state to viability under this intervention library. Early intervention preserves the reachable safe set; late intervention does not.}
\label{fig:reachable}
\end{figure}

We compare three fixed protocols over $T=56$ days: Protocol~A applies a senolytic course first (days 0 to 7) then reprogramming pulses (days 14 to 42); Protocol~B reverses the order; Protocol~C administers them simultaneously. Senolysis before reprogramming produces greater restoration of epigenetic integrity ($e_{A}(T)=0.72$, $e_{B}(T)=0.61$, $e_{C}(T)=0.66$ on the mean trajectory; stochastic integration over $n=200$ runs preserves the ranking, Supplementary S5). The reason is mechanistic: reprogramming efficacy scales with $(1-s)\,r$, so clearing senescent cells first increases the effective gain of the reprogramming pulse (\Cref{fig:worked-example}). Under a finite-horizon cost with terminal penalty
\begin{equation}
  \Phi\!\bigl(x(T)\bigr) = w_{s}\,s(T)^{2} + w_{d}\,d(T)^{2} + w_{e}\,(1-e(T))^{2} + w_{r}\,(1-r(T))^{2} + w_{f}\,f(T)^{2},
  \label{eq:terminal-J}
\end{equation}
and the weights specified in Supplementary S5, the protocol costs are $J_{A}=0.184$, $J_{B}=0.257$, $J_{C}=0.231$, so that $\hat{V}(x_{0},T)=\min\{J_{A},J_{B},J_{C}\}=J_{A}$. This is a finite-protocol approximation rather than the exact Hamilton, Jacobi, and Bellman value, demonstrating how candidate schedules can be scored by the same value-function logic. Loss of controllability with intervention start age, including a biological deadline beyond which no admissible protocol returns the state to viability, is shown in \Cref{fig:reachable}.

The ordering effect is provable analytically. With state ordering $x=(s,d,e,r,f)$, the senolytic field is $g_{\mathrm{sen}}(x)=(-\beta_{s}\,s,0,0,0,0)^{\top}$ and the reprogramming field is $g_{\mathrm{rep}}(x)=(0,0,\mu_{e}(1-s)\,r,0,0)^{\top}$. Because $g_{\mathrm{rep}}$ depends on $s$ while $g_{\mathrm{sen}}$ does not, the Lie bracket is
\begin{equation}
  [g_{\mathrm{sen}},g_{\mathrm{rep}}](x) \;=\; \bigl(0,\,0,\,\beta_{s}\,\mu_{e}\,s\,r,\,0,\,0\bigr)^{\top} \ne 0 \quad\text{whenever}\quad s>0,\;r>0.
  \label{eq:liebracket-value}
\end{equation}
At the aged initial state, $\beta_{s}\,\mu_{e}\,s_{0}\,r_{0}=(0.800)(0.150)(0.15)(0.40)=0.0072$. The bracket is positive in the epigenetic-integrity coordinate because senolysis increases the effective gain of reprogramming by reducing the inhibitory factor $s$, which is the mathematical expression of the ordering effect observed in simulation. The full derivation appears in Supplementary S5.

\begin{insightbox}[Falsifiability]
Non-commutativity is not merely a metaphor for ``sequence matters.'' It is a computable property of controlled biological vector fields. If empirical measurements showed that $g_{\mathrm{rep}}$ were independent of $s$, or that senolysis did not alter the state variable suppressing reprogramming efficacy, the Lie bracket would vanish and the ordering prediction would fail. The claim is therefore falsifiable.
\end{insightbox}

Reprogramming is not a purely beneficial epigenetic field. It also raises an oncogenic-risk and dedifferentiation coordinate whose contribution to the safety cost is state-dependent, larger when senescent or damage burden is high, so that partial reprogramming can reverse sign from beneficial to harmful as baseline damage increases \citep{banito2009,mosteiro2016}. The Supplement instantiates this negative coordinate explicitly in the worked-example dynamics and cost (Supplementary S5). The same Supplement develops an optional state-space extension in which epigenetic erosion derepresses LINE1 retroelements, activating cGAS-STING and type-I interferon signaling in a positive-feedback loop with senescence and SASP, and adds nucleoside reverse-transcriptase inhibitor and SIRT6-activator intervention vectors that yield a natural three-intervention ordering problem \citep{dececco2019,simon2019,mcintyre2023,kanfi2012}; this extension is optional and is not required for the results reported here.

The vector-field statement is useful only if the field is empirically calibrated. For intervention $j$, $g_{j}(x)=\mathbb{E}[(x(t+\Delta t)-x(t))/\Delta t\mid x(t)=x,u_{j}]$ is estimated by combining perturbational datasets across scales (LINCS L1000 and CMap \citep{lamb2006,subramanian2017}, Perturb-seq and pooled CRISPR screens, organoid panels, animal intervention studies \citep{justice2019}, and clinical multi-omics \citep{fahy2019}) using neural ODEs \citep{chen2018}, Gaussian-process state-space models, or Koopman operators with control, together with conditional perturbation autoencoders \citep{lotfollahi2023}. Every estimate must carry uncertainty, and a coordinate enters as causal only if perturbing it changes future trajectories as predicted. The full data-source taxonomy, training objective, uncertainty quantification, validation criterion, and an AAV-FOXN1 thymic-regeneration safety example are given in Supplementary S5.

\section{Implementation Architecture}\label{sec:architecture}
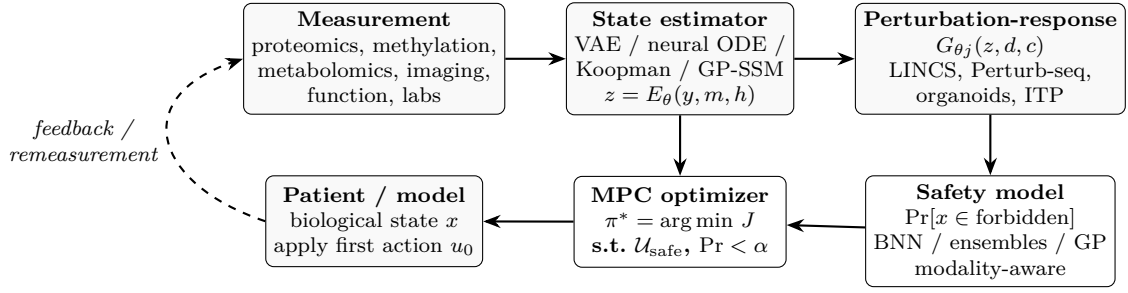
\begin{figure}[t]
\centering
\begin{tikzpicture}[
  >=Stealth,
  node distance=8mm,
  box/.style={rectangle,draw=navy,fill=gray!5,rounded corners=3pt,minimum height=12mm,minimum width=28mm,align=center,font=\scriptsize,inner sep=3pt},
  acc/.style={rectangle,draw=tealAccent,fill=tealBg,rounded corners=3pt,minimum height=12mm,minimum width=28mm,align=center,font=\scriptsize\bfseries,inner sep=3pt},
  safety/.style={rectangle,draw=amberAccent,fill=amberBg,rounded corners=3pt,minimum height=12mm,minimum width=28mm,align=center,font=\scriptsize,inner sep=3pt},
  arrow/.style={->,thick,draw=navy}
]
\node[box] (meas) {\textbf{Measurement}\\proteomics, methylation,\\metabolomics, imaging,\\function, labs};
\node[box,right=of meas] (est) {\textbf{State estimator}\\VAE / neural ODE /\\Koopman / GP-SSM\\$z=E_{\theta}(y,m,h)$};
\node[box,right=of est] (pr) {\textbf{Perturbation-response}\\$G_{\theta j}(z,d,c)$\\LINCS, Perturb-seq,\\organoids, ITP};
\node[safety,below=of pr] (safe) {\textbf{Safety model}\\$\Pr[x\in\text{forbidden}]$\\BNN / ensembles / GP\\modality-aware};
\node[acc,below=of est] (mpc) {\textbf{MPC optimizer}\\$\pi^{*}=\argmin\,J$\\s.t.\ $\Vsafe$, $\Pr<\alpha$};
\node[box,below=of meas] (patient) {\textbf{Patient / model}\\biological state $x$\\apply first action $u_{0}$};

\draw[arrow] (meas) -- (est);
\draw[arrow] (est) -- (pr);
\draw[arrow] (pr) -- (safe);
\draw[arrow] (safe) -- (mpc);
\draw[arrow] (est) -- (mpc);
\draw[arrow] (mpc) -- (patient);
\draw[arrow,dashed] (patient.west) .. controls +(-1.5,0.5) and +(-1.5,-0.5) .. (meas.west)
   node[midway,left,font=\scriptsize\itshape,align=center] {feedback /\\remeasurement};
\end{tikzpicture}
\caption{Closed-loop implementation architecture: measurement $\to$ latent-state estimation $\to$ perturbation-response and safety models $\to$ safety-constrained MPC $\to$ first action applied to the patient or model $\to$ remeasurement. The safety model gates the admissible set at every optimization step.}
\label{fig:architecture}
\end{figure}

The framework can be implemented as a closed-loop computational system that estimates a latent biological state, learns state-dependent intervention vector fields, computes safe reachable sets, selects interventions under uncertainty, and updates the model after repeated measurement. The pipeline consists of six layers (\Cref{fig:architecture}): a multimodal \textbf{measurement} layer (plasma proteomics, DNA methylation, untargeted metabolomics, immune profiling, single-cell omics, functional tests, imaging, and clinical chemistry, with rigorous preprocessing and uncertainty propagation); a \textbf{latent-state estimator} $z(t)=E_{\theta}(y(t),m(t),h(t))$ implemented as a multimodal VAE, neural ODE \citep{chen2018}, Koopman operator, or Gaussian-process state-space model, biologically anchored and trained to predict future trajectories rather than only age; a \textbf{perturbation-response learner} that estimates conditional vector fields $G_{\theta j}(z,d,c)$; a modality-aware \textbf{safety model} that estimates the probability of entering forbidden regions over short and long horizons; a \textbf{safety-constrained model-predictive controller}
\begin{equation}
  \pi^{*} \;=\; \argmin_{\pi}\;\mathbb{E}\!\left[\,\int_{t}^{t+T}\!L\bigl(z(s),u(s),s\bigr)\,ds + \Phi\!\bigl(z(t+T)\bigr)\right],
  \label{eq:mpc}
\end{equation}
subject to $u(s)\in\Vsafe(z(s),h)$ and $\Pr[z(s)\text{ forbidden}]<\alpha$, which applies only the first action before remeasurement and may recommend no treatment; and a \textbf{feedback loop} that updates the latent state via Bayesian filtering after quarterly lightweight and annual deep phenotyping. Robust MPC (conditional value at risk, worst-case objectives) is used when model uncertainty is high, and the problem becomes a POMDP over belief states under partial observation. Full layer-by-layer specifications and a two-year minimal viable implementation focused on immune aging appear in Supplementary S6.

\section{Translational Case Studies}\label{sec:case-studies}
The framework becomes useful only when instantiated in specific tissues with measurable states, admissible interventions, safety constraints, and go/no-go criteria. Three case studies, developed in full in Supplementary S7, illustrate this. In \textbf{aged thymus and immune rejuvenation}, the latent state collects thymic volume, epithelial and stromal integrity, adipose replacement, na\"ive T-cell output, repertoire diversity, inflammatory tone, and clonal-expansion risk; a staged policy begins with reversible metabolic and inflammatory optimization, proceeds to transient thymopoietic stimulation (IL-7 axis, GH secretagogues with metformin co-administration), and considers durable FOXN1 programs only at a higher evidence threshold, with success defined by improved na\"ive output, repertoire diversity, and vaccine response without autoimmunity or clonal expansion. In \textbf{sarcopenia}, the state separates mass from quality, mitochondrial function, neuromuscular-junction integrity, fibrosis, adiposity, and functional performance, so that the controller penalizes mass gain without strength, power, or metabolic improvement; reversible interventions (resistance exercise, protein and amino-acid optimization, urolithin~A and mitophagy-targeting agents \citep{rajman2018}, rapamycin schedules) precede biologics (myostatin/activin inhibitors) and durable follistatin gene therapy. In \textbf{ovarian aging}, follicular reserve is a dominant irreversible depletion coordinate ($dF/dt\le 0$ under the standard small-molecule set), making the tissue a stringent test of biological deadlines, with the control goal being maintenance of fertility potential over a defined horizon under strict state-dependent admissibility around conception and pregnancy. Each case specifies measurement layers, candidate interventions, safety constraints, and go/no-go criteria, and each tests whether control-value reduction predicts functional improvement better than Hallmark labels or clock reversal alone.

\section{Validation and Power Analysis}\label{sec:validation}
A control-theoretic framework should be judged by whether it improves drug-discovery decisions, and AI-driven discovery pipelines \citep{zhavoronkov2019a,zhavoronkov2019b,partridge2020} provide a natural validation environment. The identification of dual-purpose targets implicated in both aging hallmarks and age-associated diseases has been demonstrated on AI platforms \citep{pun2022dualpurpose}, and programs since 2021 against six aging-associated targets (TNIK, PHD1/2, QPCTL, NLRP3, NR3C1, TEAD) constitute a retrospective testbed in which each program is a decision sequence amenable to comparison of target-scoring frameworks. For each candidate one can compute a Hallmark score, age-association score, disease-association score, network-centrality score, druggability score, toxicity-risk score, and \textbf{control-value score} (estimated reduction in restoration cost $V$). The central hypothesis is that control-value score outperforms Hallmark score for predicting translational advancement, functional efficacy, and toxicity-adjusted therapeutic index. An internal pilot is statistically limited: detecting a Spearman correlation near 0.45 with 80\% power at two-sided $\alpha=0.05$ requires roughly 36 to 40 compounds, and advancement is influenced by potency, pharmacokinetics, intellectual property, and strategic decisions beyond biological efficacy. A stronger program expands to public corpora (DrugAge, the Interventions Testing Program, known geroprotectors, and matched negative controls), is pre-registered with frozen compound list, model version, intervention library, cost function, and endpoints, and prefers prospective scoring in aged organoid panels. The full design, the four-quadrant Hallmark-versus-control-value prediction, the pre-registration template, and explicit failure criteria appear in Supplementary S8. The framework fails if control-value scoring does not outperform Hallmark or network scoring, if state-dependent predictions do not improve outcomes, or if non-commutativity, irreversibility, or safety-constraint predictions fail when tested; none of these failure modes, and none of the corresponding successes, has yet been demonstrated in pre-registered, independent, adequately powered studies.

\section{Twenty Falsifiable Predictions}\label{sec:predictions}
The following twenty predictions distinguish the control-theoretic framework from Hallmark enumeration, biomarker optimization, and single-mechanism theories. Each is a testable, falsifiable hypothesis; full statements, tests, and falsification conditions are given in Supplementary S9. They group into state-dependent efficacy (1 through 4), order and combination (5 through 8), controllability and irreversibility (9 through 12), target prioritization (13 through 16), and systems-level behavior (17 through 20).

\begin{enumerate}\setlength\itemsep{1pt}
  \item Sign and magnitude of effect depend on baseline state beyond chronological age.
  \item Non-responders to rapamycin, senolytics, or partial reprogramming are identifiable from pre-treatment latent state.
  \item At least one major intervention has a measurable state region where it is harmful despite being beneficial on average.
  \item Optimal dose depends on state, not a single target-engagement curve.
  \item In fibrotic/inflamed aged tissues, senolysis before reprogramming restores function better than the reverse order.
  \item In fibrotic tissues, antifibrotic remodeling before regenerative stimulation outperforms the reverse order.
  \item The magnitude of empirical non-commutativity correlates with the estimated Lie bracket of intervention vector fields.
  \item Rationally chosen low-dose combinations match or exceed high-dose monotherapy at lower toxicity.
  \item Structurally aged tissues have measurable controllability boundaries beyond which no admissible small molecule restores function even if clocks improve.
  \item Recovery dynamics after standardized perturbation predict future decline better than static biomarkers.
  \item Ovarian reserve and hippocampal neuronal loss have identifiable deadlines beyond which standard interventions cannot restore function.
  \item State-space variance and autocorrelation increase measurably before clinical frailty transitions.
  \item Some high-control-value targets lie outside canonical Hallmark annotations.
  \item Control-value reduction predicts advancement better than Hallmark membership alone.
  \item High network-centrality targets without high control-value reduction underperform in translation.
  \item Modality-matched targeting outperforms target-only selection.
  \item Closed-loop adaptive MPC outperforms fixed-schedule protocols at matched cumulative exposure.
  \item Reversibility-hierarchy escalation reduces serious adverse events without sacrificing efficacy.
  \item Controllability structure differs between sexes (ovarian dominance, distinct immune kinetics, different mTOR/IGF-1 sensitivities).
  \item Interventions targeting shared aging mechanisms reduce multi-system disease incidence more efficiently than single-disease drugs at matched cost and safety.
\end{enumerate}

Each prediction derives from a specific formal object of the framework rather than from biological intuition alone. \Cref{tab:pred-mapping} makes this mapping explicit for the five predictions that are both riskiest and most directly tied to the formalism; the full mapping for all twenty predictions appears in Supplementary S9. A prediction is riskiest when it is independently testable and is not a built-in modeling assumption, so that its failure would refute the corresponding formal object.

\begin{table}[ht]
\centering
\footnotesize
\caption{Mapping of the five riskiest predictions to the formal object each derives from. Failure of the prediction refutes the associated formal claim.}\label{tab:pred-mapping}
\begin{tabularx}{\textwidth}{@{}c X l@{}}
\toprule
\textbf{Pred.} & \textbf{Statement (abbreviated)} & \textbf{Formal object}\\
\midrule
1  & The same intervention is beneficial in some states and harmful in others & Safety-cost sign: $g_{j}^{\top}\nabla V$ changes sign\\
2  & Sequenced interventions with nonzero Lie bracket depend on order & Lie bracket $[g_{A},g_{B}]\ne 0$ (ordering)\\
3  & Each tissue has a loss level beyond which no admissible intervention restores function & Viability-kernel boundary / non-responder state\\
5  & Targets ranked by control value outperform Hallmark-count ranking & Value-function gradient identifies high-leverage node\\
17 & Control-value biological age predicts hazard better than chronological age & $\mathrm{BA}_{c}$ as monotone map of $V(x)$\\
\bottomrule
\end{tabularx}
\end{table}

\section{Empirical Application: Scoring Interventions Across Biological Epochs}\label{sec:compound-scoring}
To show that the framework generates actionable, quantitative predictions, we applied an epoch-stratified scoring rubric to a curated set of interventions drawn from DrugAge (3{,}423 entries), the NIA Interventions Testing Program (63 entries across $\sim$50 compounds), ClinicalTrials.gov (638 aging/longevity trials), and OpenTargets (2{,}559 aging-associated targets). Each intervention was mapped onto a 20-dimensional biological state space (mTOR, AMPK, sirtuin, inflammatory tone, senescent burden, epigenetic integrity, proteostasis, mitochondrial function, NAD$^+$, stem-cell reserve, telomere length, immune function, vascular integrity, fibrosis, metabolic flexibility, hormonal regulation, circadian amplitude, microbiome diversity, neuroplasticity, genomic stability) with a vector field $g_i\in\mathbb{R}^{20}$, and scored on eight axes (control leverage, safety margin, controllability expansion, durability, reversibility, combinability, evidence tier, sequence position) at each of five biological epochs. The composite score at epoch $k$ is $\text{Composite}_k = 100\times\sum_{i=1}^{8} w_{k,i}\cdot\text{Axis}_i$ with epoch-specific weights, and the Lifetime Integrated Score (LIS) is $\sum_{k=1}^{5}\text{Composite}_k$. The full rubric, axis definitions, epoch weights, and the complete scored dataset (Supplementary Table~S1) are given in Supplementary S10; \Cref{tab:top20-lis} reports the top twenty interventions by LIS.

\begin{table}[ht]
\centering
\caption{Top 20 aging interventions by Lifetime Integrated Score (LIS). Evidence tier: 1 = ITP/Phase~3 validated; 2 = Phase~1 or 2 / strong preclinical; 3 = preclinical only.}\label{tab:top20-lis}
\small
\begin{tabular}{clcccc}
\toprule
\textbf{Rank} & \textbf{Intervention} & \textbf{Tier} & \textbf{Peak} & \textbf{Best Ep.} & \textbf{LIS} \\
\midrule
    1 & Caloric Restriction (30\%) & 1 & 72.1 & 1 & 320.2 \\
    2 & Aerobic Exercise ($\geq$150 min/wk) & 1 & 70.9 & 2 & 316.7 \\
    3 & Dapagliflozin & 1 & 69.3 & 1 & 309.3 \\
    4 & Empagliflozin & 1 & 68.2 & 1 & 306.0 \\
    5 & Resistance Exercise (2 to 3$\times$/wk) & 1 & 67.6 & 1 & 301.5 \\
    6 & Liraglutide & 1 & 67.2 & 1 & 298.9 \\
    7 & Intermittent Fasting (16:8) & 2 & 69.1 & 1 & 298.2 \\
    8 & Telmisartan & 1 & 65.7 & 1 & 289.4 \\
    9 & Semaglutide & 1 & 63.8 & 2 & 286.6 \\
    10 & Sulforaphane & 2 & 65.2 & 1 & 285.6 \\
    11 & Berberine & 2 & 64.5 & 1 & 281.7 \\
    12 & Urolithin~A & 2 & 62.9 & 1 & 279.3 \\
    13 & Losartan & 1 & 63.7 & 1 & 278.4 \\
    14 & Methionine Restriction & 3 & 64.5 & 1 & 278.4 \\
    15 & Time-Restricted Eating & 2 & 63.9 & 1 & 277.6 \\
    16 & Pioglitazone & 1 & 62.6 & 1 & 276.2 \\
    17 & Lithium (low dose) & 2 & 63.0 & 1 & 275.4 \\
    18 & Fenofibrate & 1 & 61.4 & 1 & 271.3 \\
    19 & Pterostilbene & 3 & 61.9 & 1 & 268.3 \\
    20 & Magnesium supplementation & 1 & 60.6 & 1 & 266.2 \\
\bottomrule
\end{tabular}
\end{table}

Several observations emerge. Lifestyle interventions dominate the top ranks: caloric restriction (LIS 320.2) and aerobic exercise (316.7) outperform all pharmacological agents through broad vector-field coverage, high safety margins, and Tier~1 evidence. SGLT2 inhibitors emerge as the highest-ranked pharmacological class (dapagliflozin 309.3, empagliflozin 306.0), outranking rapamycin, senolytics, and NAD$^+$ precursors through multi-dimensional effects and Tier~1 cardiovascular-outcome evidence, a prediction testable in head-to-head geroscience trials, with GLP-1 receptor agonists ranking third (liraglutide 298.9, semaglutide 286.6). All scores decline monotonically with epoch, formalizing the intuition that prevention is more efficient than late intervention through narrowing safety envelopes and diminishing substrate. Rapamycin ranks outside the top fifteen despite gold-standard ITP lifespan evidence, because its immunosuppressive safety penalty and narrow vector field reduce its composite score relative to multi-dimensional agents; the framework predicts intermittent low-dose deployment at Epoch~II to III rather than chronic administration. A central implication is that the current intervention library is incomplete, providing meaningful control over roughly 20\% of the state space ($C_0\approx 0.20$), which motivates an iterative control cycle mapping control gaps, directing discovery toward uncontrollable dimensions, and re-estimating controllability with the expanded library, so that the framework is intended to operate as an iterative pipeline for rational gerotherapeutic prioritization rather than a one-shot solution; the automated learning of vector fields that would close this loop is future and companion work.

\section{Relationship to Existing Frameworks and Formal Methods}\label{sec:existing}
The control-theoretic framework complements, rather than replaces, existing aging theories; each supplies content the control layer depends on. Hallmark annotations map onto state coordinates; damage inventories inform terminal cost; SENS repair categories map onto intervention vector fields; information-loss theories motivate specific reversible coordinates; geroscience provides clinical context; hyperfunction theory supplies a key vector field whose state-dependent sign is made explicit.

The framework also stands in a specific relation to the resilience and critical-transition literature. Work on early-warning signals for critical transitions \citep{scheffer2009,scheffer2012} and on universal resilience patterns in complex networks \citep{gao2016} shows that systems approaching a tipping point exhibit critical slowing down, rising variance, and reduced recovery rate before overt collapse. These are descriptive, intervention-free diagnostics: they detect that a system is losing dynamical resilience but do not specify which input restores it. The control-cost formulation adds a prescriptive layer on top of this descriptive picture. Loss of resilience appears here as declining conditioning of the controllability Gramian and rising restoration cost $V$, and the same slowing of recovery that serves as an early-warning signal is, in control terms, the quantity to be reversed. Beyond detection, the value function specifies an intervention-relative, safety-constrained objective: which admissible control most reduces $V$ from the current state, at what risk, and within which reachable safe set. Critical slowing down thus becomes an input to the controller rather than an endpoint of the analysis.

The operational difference from the Hallmarks scheme can be stated concretely. The Hallmarks organize mechanisms into categories and support the inference that a mechanism is a candidate target, but they do not by construction order interventions, identify who will not respond, or predict when a beneficial intervention becomes harmful. The control formulation yields three operationally distinct outputs. First, explicit intervention ordering: when two intervention vector fields have a nonzero Lie bracket, the framework predicts which sequence reduces $V$ more, a statement absent from any hallmark annotation. Second, a predicted non-responder state: an initial state lying outside the viability kernel for a given library cannot be restored by any admissible protocol in that library, identifying non-responders from the pre-treatment state rather than post hoc. Third, a prospectively testable state-dependent sign reversal: the sign of an intervention is set by the projection of its vector field onto the gradient of $V$, so the framework predicts a state region in which a normally beneficial intervention becomes harmful. These three outputs are the content the hallmark ontology does not supply. The recurring gap the control layer fills is the absence of a state vector, dynamics, ordering, an admissible set, and a continuous safe value function in place of binary controllability or annotation-based ranking; a full cross-framework comparison is tabulated in Supplementary S10. The formulation also interfaces with three adjacent formal methods, developed in Supplementary S10: \emph{hybrid automata} \citep{henzinger2000,olde_loohuis2014,olde_loohuis2012} represent discrete clinical stages (robust, pre-frail, frail, disabled, terminal) with mode-dependent flows and enable model-checking of safety properties; \emph{asymmetric signaling games} \citep{massey_mishra2018,casey2020} model cell-cell communication under selection, where pooling equilibria underlie inflammaging and immune evasion and interventions act on equilibrium structure; and \emph{anti-hallmarks and synthetic lethality} \citep{bryant2005,brennan2008} predict that high-value targets lie outside canonical Hallmark vocabulary in dependencies the aged state acquired through compensatory remodeling, consistent with the high ranking of SGLT2 inhibitors and GLP-1 receptor agonists above.

\section{Limitations}\label{sec:limitations}
The framework has substantial limitations. The latent state is not observable, and inferring it from measurements is an identifiability problem compounded when interventions perturb the measurement mapping itself; it should be addressed through multimodal over-determination, perturbational experiments, biological priors, model comparison on held-out perturbational data, and sensitivity analysis. Intervention data are sparse and biased toward cancer cell lines, young mice, and healthy volunteers, so extrapolation to aged, frail, and comorbid populations is risky and prospective aging-specific datasets are essential. The full stochastic control problem is intractable in closed form, and the linearizations, short-horizon MPC, reinforcement learning, and mechanistic submodels used in practice introduce errors that must be quantified. Long-horizon safety prediction is weak because aging drug programs are young and relevant outcomes have long latencies; the framework keeps safety uncertainty in the optimization and prefers reversible modalities when uncertainty is high. Value-function weights reflect patient preferences, clinical judgment, and societal values and must be elicited rather than derived from biology, and control models must be shared with frozen parameters, code, and pre-specified endpoints to be falsifiable. Finally, the framework is a research program: it does not authorize any specific clinical protocol or off-label use, which must follow regulatory, ethical, and evidentiary standards independent of the framework. A five-phase scalability roadmap, from single-organ animal and organoid models through human pilots, multi-organ combination validation, durable-modality integration, and population-scale geroscience, with defined deliverables and failure modes at each phase, is given in Supplementary S6.

\section{Discussion}\label{sec:discussion}
The core claims are that aging is progressive loss of safe controllability; biological age is the minimum safe control cost of restoring or maintaining function; drugs are state-dependent vector fields; combinations and sequences matter through Lie-bracket structure; safety is a first-class constraint that defines the admissible control set; and modality class determines the reversibility hierarchy of controls. The distinguishing contribution is not the general idea that drugs perturb biological systems, which is standard systems pharmacology, but the casting of gerotherapeutic discovery as a constrained stochastic control problem with explicit state space, safety constraints, irreversibility coordinates, and an intervention-relative definition of biological age. This yields specific, testable consequences: responder stratification by state, order-dependent combinations, controllability boundaries, modality-matched escalation, adaptive MPC, and safety-constraint modeling with forbidden regions and reward-hacking prevention. It addresses the gap between descriptive biology and translational decision-making by asking, at each step, which admissible intervention most reduces the minimum safe cost of functional restoration given the current state estimate and its uncertainty. Several caveats hold: identifiability, data bias, computational complexity, and weak safety modeling are serious; parameters must be empirically calibrated and predictions pre-registered; and value-function weights cannot be derived from biology alone. The central empirical test is whether control-value reduction predicts translational success better than alternative scoring methods, conductable retrospectively on existing portfolios, prospectively in aged organoid and animal panels, and adaptively in clinical cohorts.

\section{Materials and Methods}\label{sec:methods}
All simulations in the worked example (\Cref{sec:worked-example}) used the specified stochastic differential equation system. Deterministic mean-trajectory integration used explicit Euler stepping with $\Delta t=0.25$ days over a 56-day horizon, with state variables clamped to $[0,1]$ after each step; the stochastic reference integration used the same step size with additive Wiener increments scaled by the specified diagonal noise matrix. Control functions were normalized so that a senolytic or TNIK course and a reprogramming pulse each integrate to unity. The value-function approximation is over three hand-specified policies and does not solve the full Hamilton, Jacobi, and Bellman equation. The Lie-bracket computation was performed analytically from the vector-field specifications and independently checked in SymPy. All computations are reproducible from the equations, parameters, and initial conditions provided; a reference implementation (Python with NumPy and SciPy) will be released on publication under an open license, and no patient data are analyzed.

In accordance with the ICMJE Recommendations (January 2024 update) and the WAME recommendations on chatbots and generative AI \citep{icmje2023}, the authors disclose that large language models (Anthropic Claude Opus and Sonnet families; OpenAI GPT-4 and GPT-5 families, 2024 to 2026 releases) were used as drafting and editing tools for prose, copy-editing, reference cross-checking against author-provided sources, and LaTeX assistance, and that retrieval-augmented assistants located candidate citations subsequently verified against the primary literature. AI systems did not originate the scientific framework, did not generate claims or predictions not specified by the human authors, did not produce any citation accepted without human verification, and did not carry out the numerical simulations or analytic derivations. AI systems are not listed as authors. The human authors reviewed, edited, and approved every sentence, equation, figure, table, reference, and numerical claim, and accept full responsibility for the accuracy, originality, and integrity of the work. The manuscript was checked for textual overlap against prior publications and indexed sources; all retained passages are original or properly quoted and cited; the submission contains no fabricated or falsified data and no unverified AI-generated citations. No human or animal experiments were performed for this manuscript; translational case studies outline prospective designs requiring institutional review, informed consent, and regulatory oversight. The authors disclose affiliations with Insilico Medicine and collaborating institutions; specific financial disclosures appear in the journal submission record. The framework is proposed as an open scientific program requiring independent replication beyond any single organization.

\section{Conclusion}\label{sec:conclusion}
Aging theory has matured from evolutionary reasoning through damage catalogues, Hallmark ontology, information models, geroscience, hyperfunction, and network and resilience frameworks. Each has clarified part of the biology; none, by itself, specifies how to act in a measured biological state. We have proposed a control-theoretic framework designed to supply that layer: a latent biological state, a functional viability set, stochastic aging dynamics, intervention vector fields, a minimum-safe-cost definition of biological age, explicit irreversibility coordinates and absorbing boundaries, modality-aware admissible sets, and safety as a first-class constraint. A worked aged-liver example with an analytic Lie bracket and protocol scoring is explicit enough to be reproduced and falsified; an implementation architecture covers measurement, latent-state estimation, perturbation-response learning, safety modeling, constrained model-predictive control, and closed-loop updating; three translational case studies specify state vectors, intervention libraries, constraints, and go/no-go criteria; a validation plan with power analysis and external replication is provided; and twenty pre-registered, falsifiable predictions are enumerated. The framework will succeed or fail empirically. If control-value scoring does not outperform Hallmark, clock, or network baselines, and if order-dependence, controllability boundaries, modality effects, and adaptive-control advantages fail to appear, it will be rejected. The framework supplies aging drug discovery with equations of motion, admissible sets, value functions, reversibility hierarchies, and falsifiable predictions, and its claims are stated so that adequately powered, pre-registered studies can confirm or refute them. The present paper establishes the framework and its predictions; the deeper mathematical and engineering development, including the full geometric-control treatment of noncommutativity and reachability, formal identifiability and estimation of the vector fields, and hybrid-automaton and stochastic-transition models of irreversibility, is reserved for a companion sequel.

\section*{Acknowledgements}
The authors thank Charles Cantor for his input and support. The authors thank colleagues at Insilico Medicine and collaborating institutions for valuable discussions.

\bibliographystyle{unsrtnat}
\bibliography{references}

\end{document}